\documentclass{PoS}
\usepackage{graphicx}
\usepackage{amsmath}
\usepackage{amssymb}

\title{
   \begin{picture}(0,0)(0,0)%
   \put(350,75){\makebox(0,0)[l]{\textnormal{\normalsize KEK-CP-331}}}%
   \end{picture}%
   Nucleon axial and tensor charges with dynamical overlap quarks
}

\ShortTitle{Nucleon axial and tensor charges with dynamical overlap quarks}

\author{
JLQCD collaboration:\,\, 
\speaker{N.~Yamanaka}$^a$, 
H.~Ohki$^{b}$, 
S.~Hashimoto$^{c,d}$
and
T.~Kaneko$^{c,d}$,
\\
\llap{$^a$}iTHES Research Group, RIKEN,
           2-1 Hirosawa, 351-0115 Saitama, Japan\\
\llap{$^b$}RIKEN BNL Research Center, 
  Brookhaven National Laboratory, Upton NY 11973, USA\\
\llap{$^c$}High Energy Accelerator Research Organization (KEK), 
           Tsukuba 305-0801, Japan\\
\llap{$^d$}School of High Energy Accelerator Science, 
           The Graduate University for Advanced Studies\\
           (Sokendai), Tsukuba 305-0801, Japan\\
E-mail: \email{nodoka.yamanaka@riken.jp}
}

\abstract{
We report on our calculation of the nucleon axial and tensor charges
in 2+1-flavor QCD with dynamical overlap quarks.
Gauge ensembles are generated at a single lattice spacing 0.12 fm
and at a strange quark mass close to its physical value. 
We employ the all-mode-averaging technique to 
calculate the relevant nucleon correlation functions,
and the disconnected quark loop is efficiently calculated 
by using the all-to-all quark propagator.
We present our preliminary results 
 for the isoscalar and isovector charges
obtained at pion masses $m_\pi\!=\!450$ and 540~MeV.
}

\FullConference{The 33rd International Symposium on Lattice Field Theory\\
		14 -18 July 2015\\
		Kobe International Conference Center, Kobe, Japan*}

\begin{document}

\section{Introduction}

The nucleon charges represent the nonperturbative nature of QCD, and are also relevant to
 the search for new physics beyond the Standard Model.
The nucleon axial charge $\Delta q$ for the quark flavor $q$ is defined by
\begin{eqnarray}
\langle N (p,s) | \bar q \gamma_\mu \gamma_5 q | N (p,s) \rangle
&=& 
2 m_N s_\mu \Delta q,
\end{eqnarray}
where $p$ and $s$ are four-vector momentum and polarization of the nucleon, 
respectively.
This probes the quark contribution to the nucleon spin,
and is a fundamental quantity to understand the so-called 
``proton spin puzzle''. 
The tensor charge $\delta q$ is defined by
\begin{eqnarray}
\langle N (p,s) | \bar q i \sigma_{\mu \nu} \gamma_5 q | N (p,s) \rangle
&=& 
2 (s_\mu p_\nu - s_\nu p_\mu ) \delta q
,
\end{eqnarray}
and describes the contribution of possible tensor-type interactions
beyond the Standard Model to nucleon observables.
It appears in the search for new physics 
through precision measurements of the electric dipole moment and 
the $\beta$ decays.

We recently calculated the strange-quark scalar charge
$\langle N|\bar{s}s|N\rangle$
in lattice QCD with dynamical overlap quarks.
In Ref.~\cite{ohki}, we utilized the Feynman-Hellmann theorem to obtain the scalar charge,
whereas we directly calculated 
the nucleon disconnected three-point function 
by using the all-to-all quark propagator 
\cite{takeda,ohki2}.
In this article, we extend the latter study to the axial and tensor charges.
We report our preliminary results for the isovector charges, 
$g_A\!=\!\Delta u-\Delta d$ and $g_T\!=\!\delta u-\delta d$,
as well as those for the isoscalar charges
$g_A^s\!=\!\Delta u+\Delta d$ and $g_T^s\!=\!\delta u+\delta d$,
and the strange-quark charges, $\Delta s$ and $\delta s$,
which receive contributions from the disconnected diagram.

\section{Simulation method}

We simulate three-flavor QCD 
using the Iwasaki gauge action and overlap quark action.
Numerical simulations are remarkably accelerated 
by simulating trivial topological sector with a modification 
of the gauge action~\cite{exW}.
Gauge ensembles are generated on a $16^3\!\times\!48$ lattice 
at a lattice spacing $a\!=\!0.12$~fm and 
with a strange quark mass $m_s\!=\!0.080$
close to its physical value $m_{s,\rm phys}\!=\!0.081$.
In this article, we present results 
at two values of degenerate up and down quark masses,
$m_{ud}\!=\!0.035$ and 0.050,
which correspond to the pion masses 
$m_\pi\!\sim\!450$ and 540 MeV, respectively.
We note that simulations at lighter pion masses
290\,--\,380~MeV are in progress.

We calculate the nucleon three-point function
\begin{eqnarray}
C_{{\rm 3pt}} ( t_{\rm src} ,   {\bf y}_{\rm src}, \Delta t,  \Delta t^\prime)
&=&
\frac{1}{N_s^6} \sum_{{\bf x} ,{\bf z} }
\Biggl\{
{\rm tr}_s 
\Bigl[
P 
\bigl<
N ({\bf x } , t_{\rm src}+\Delta t^\prime ) 
{\mathcal O}_\Gamma ({\bf z} , t_{\rm src}+\Delta t) \bar N ( {\bf y}_{\rm src} , t_{\rm src} )
\bigr>
\Bigr]
\nonumber\\
&& \hspace{3em}
-
\bigl<
{\mathcal O}_\Gamma ({\bf z} , t_{\rm src}+\Delta t)
\bigr>
{\rm tr}_s 
\Bigl[
P 
\bigl<
N ({\bf x } , t_{\rm src}+\Delta t^\prime) \bar N ( {\bf y}_{\rm src} , t_{\rm src} )
\bigr>
\Bigr]
\Biggr\}
,
\label{eq:3ptfunction}
\end{eqnarray}
where $P = (1+\gamma_4) \gamma_5 \gamma_3$
is the polarization matrix, and use the quark bilinear operator 
${\mathcal O}_\Gamma\!=\!\bar{q}\gamma_3 \gamma_5 q$ 
for the axial charge ($\Gamma\!=\!A$) and 
$i\bar{q}\sigma_{03}\gamma_5 q$ for the tensor charge ($\Gamma=T$), 
respectively.
The nucleon interpolating operator is 
$N = \epsilon_{abc} (u_a^T C \gamma_5 d_b) u_c$, 
for which we apply the Gaussian smearing 
$ q ({\bf x } , t) =
\sum_{\bf y } 
\left\{ 
   ( 1+\omega H/4N )^N
\right\}_{{\bf x } ,{\bf y }}
q_{\rm local} ({\bf y } , t)
$
with 
$H_{{\bf x},{\bf y}} = \sum_{i=1}^3 
(\delta_{{\bf x},{\bf y}-\hat{\bf i}} + \delta_{{\bf x},{\bf y}+\hat{\bf i}})$
in order to enhance the overlap with the nucleon ground state.
The parameters $\omega\!=\!20$ and $N\!=\!400$ are chosen 
from our experience in Refs.~\cite{takeda,ohki2}.

The nucleon charges are extracted from the ratio
\begin{equation}
R(t)=
Z_\Gamma \frac{C_{\rm 3pt}( \Delta t, \Delta t^\prime)}{C_{\rm 2pt}( \Delta t^\prime)}
\xrightarrow[\Delta t,\Delta t^\prime - \Delta t \to \infty\\ ]{ } 
\frac{\langle N | Z_\Gamma {\mathcal O}_\Gamma | N \rangle}{2m_N},
\label{eq:nucleoncharge}
\end{equation}
where $C_{\rm 2pt}$ is the nucleon two-point function with the same nucleon interpolating fields and the same time separation $\Delta t'$ as those for the three-point function.
The arguments $(t_{\rm src},{\bf y}_{\rm src})$ of the correlators are omitted 
(see the following discussion),
and $Z_\Gamma$ is the renormalization factor in 
the $\overline{\rm MS}$ scheme at $\mu\!=\!2$~GeV.
In this preliminary analysis,
we use the values in Ref.~\cite{noaki},
which are for the flavor non-singlet bilinear operators, 
both for the isovector and isoscalar charges, ignoring possible shift due to the presence of the  disconnected contribution.

We calculate the quark loop in the disconnected diagram 
by using the all-to-all quark propagator~\cite{A2A:SESAM,A2A:TrinLat}.
Namely, the propagator is decomposed into the contribution of 
the low-lying modes of the overlap-Dirac operator $D$
\begin{equation}
(D^{-1} )_{\rm low}  (x,y)
=
\sum_{i=1}^{N_e}
\frac{1}{\lambda^{(i)}}
v^{(i)} (x) v^{(i) \dagger} (y) 
\label{eq:lowmode}
\end{equation}
and the remaining part $(D^{-1} )_{\rm high}$.
Here $\lambda^{(i)}$ and $v^{(i)}$ represent
the $i$-th lowest eigenvalue of $D$ and the associated eigenvector, 
respectively. The number of low-modes is set to $N_e\!=\!160$. 

The contribution of the remaining high-modes is estimated 
by the noise method~\cite{noise}.  
For each configuration, we prepare a complex $Z_2$ noise vector $\eta (x)$,
which is diluted into $N_d = 3\times 4 \times N_t / 2$ vectors $\eta^{(d)} (x)$
$(d\!=\!1,\cdots,N_d)$ with respect to the color and spinor indices 
as well as the temporal coordinate. 
For more details on our implementation, 
see Refs.~\cite{takeda,ohki2}.
The high-mode contribution is then given by 
\begin{equation}
(D^{-1} )_{\rm high}  (x,y)
=
\sum_{d=1}^{N_d}
\psi^{(d)} (x) \eta^{(d) \dagger} (y) 
,
\label{eq:highmode}
\end{equation}
where $\psi^{(d)} (x)$ is 
the solution of $D \psi^{(d)} \!=\! (1-P_{\rm low}) \eta^{(d)}$
with $P_{\rm low}$ the projection operator to the eigenspace
spanned by the low-modes $\{v^{(i)}\}$.

\begin{figure}[t]
\begin{minipage}{0.5\hsize}
\begin{center}
\includegraphics[width=7.5cm]{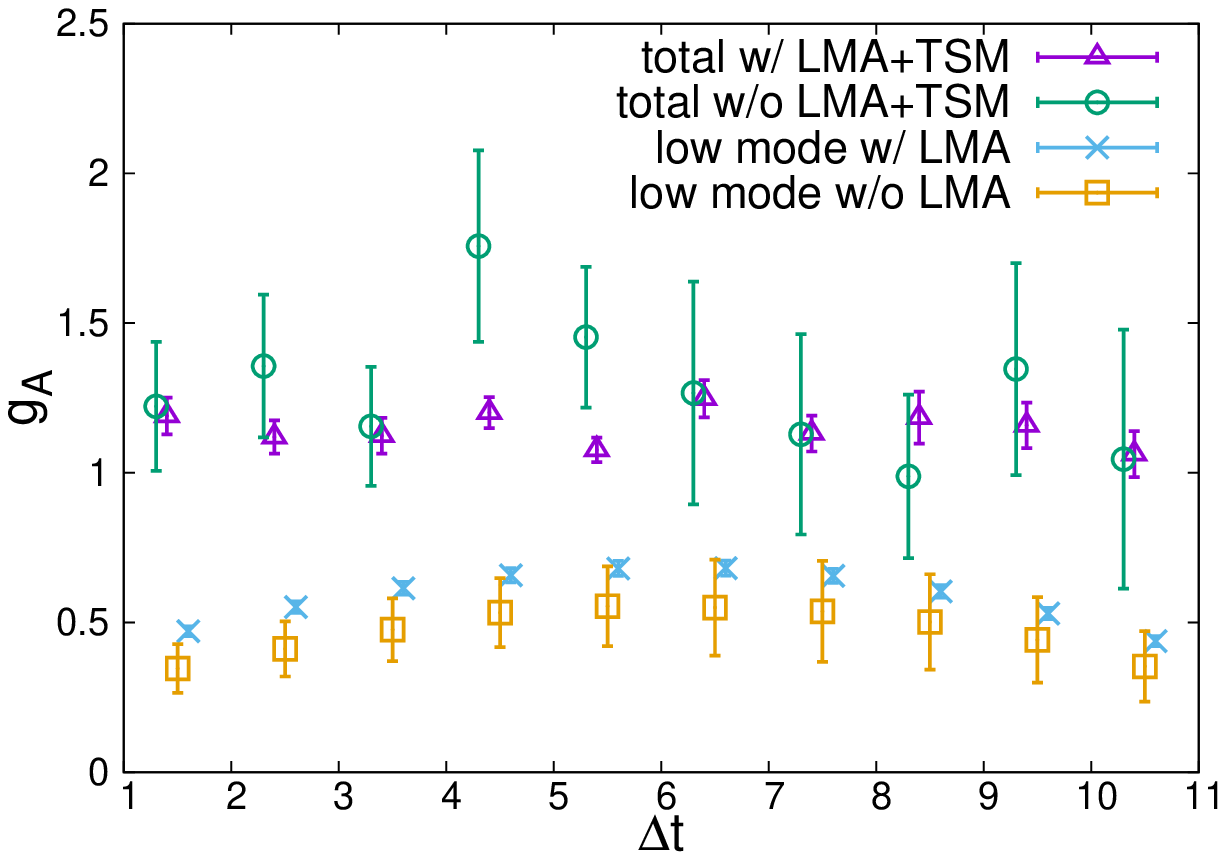}
\end{center}
\end{minipage}
\begin{minipage}{0.5\hsize}
\begin{center}
\includegraphics[width=7.5cm]{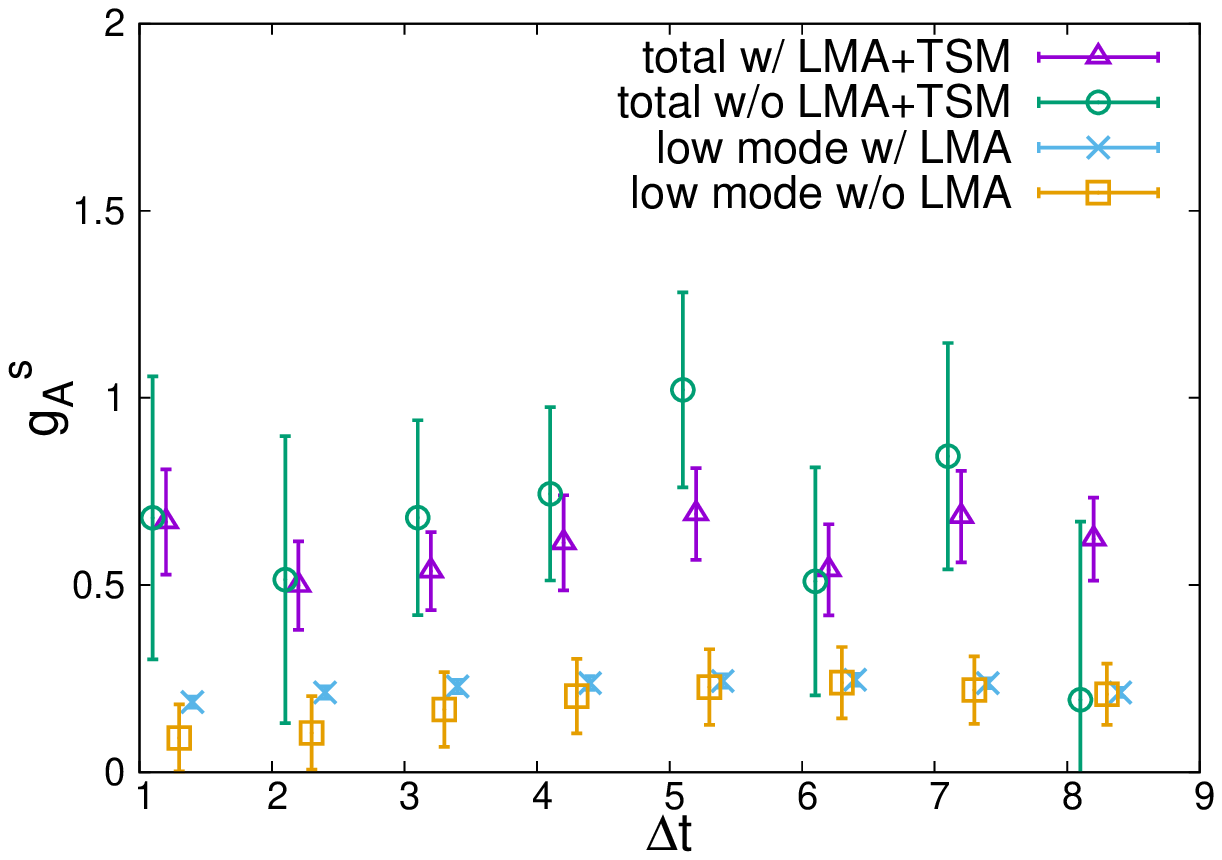}
\end{center}
\end{minipage}
\caption{\label{fig:lmaama}
Improvement of statistical accuracy in isovector (left panel) 
and isoscalar (right panel) charges. 
The circles (triangles) shows the charge calculated without (with) LMA and TSM.
We also compare the low-mode contributions to the charge,
which are calculated by using $C_{\rm 3pt,low}$ and a similar part for $C_{\rm 2pt}$ in Eq.~(\protect\ref{eq:nucleoncharge}),
before (square) and after (cross) LMA.
}
\end{figure}

We observe that the nucleon correlators 
constructed by the all-to-all propagator suffer from 
a large statistical noise,
since only single noise vector is used for each configuration.
We therefore employ the all-mode averaging technique~\cite{blumama}
to calculate $C_{\rm 2pt}$ and $C_{\rm 3pt}$.
Let us consider the decomposition 
$C_{\rm 3pt}=C_{\rm 3pt,low} + C_{\rm 3pt,high}$,
where $C_{\rm 3pt,low}$ represents the contribution 
in which the low-mode truncation (\ref{eq:lowmode}) is 
used for all the four quark propagators. 
For the remaining contribution $C_{\rm 3pt,high}$,
we use the so-called point-to-all propagator 
$\psi_{\rm pt}(x)$, 
which is obtained by solving $D\psi_{\rm pt}\!=\!b$
with $b({\bf x},t)\!\propto\!
\{(1+\omega H/4N)^N\}_{\bf x,x^\prime} \delta_{{\bf x}^\prime,{\bf y}_{\rm src}}\delta_{t,t_{\rm src}}$ with a source point $({\bf y}_{\rm src},t_{\rm src})$.

We employ the low-mode averaging (LMA)~\cite{lma:ds,lma:ghlww} 
for $C_{\rm 3pt,low}$. Namely this contribution is replaced by
that averaged over the source points $({\bf y},t_{\rm src})$.
We take one point per time-slice,
and the number of the source points is $N_{\rm src,low}\!=\!48$.
LMA in our study can be expressed as 
\begin{equation}
C_{\rm 3pt,low}\left(t_{\rm src}, {\bf y}_{\rm src}, \Delta t, \Delta t^\prime\right)
\to 
C_{\rm 3pt,low}\left(\Delta t, \Delta t^\prime\right)
=
\frac{1}{N_{\rm src,low}}
\sum_{t_{\rm src}=1}^{N_{\rm src,low}}
C_{\rm 3pt,low} \left(t_{\rm src}, {\bf y}_{\rm src}(t_{\rm src}), \Delta t, \Delta t^\prime \right)
,
\end{equation}
where ${\bf y}_{\rm src}$ is considered as a function of $t_{\rm src}$.

For the high-mode contribution $C_{\rm 3pt,high}$,
we use the truncated solver method (TSM)~\cite{tsm} and 
replace $C_{\rm 3pt,high}\left(t_{\rm src}, {\bf y}_{\rm src}, \Delta t , \Delta t^\prime\right)$
by 
\begin{eqnarray}
C_{\rm 3pt,high}\left(\Delta t, \Delta t^\prime \right)
& = &
C_{\rm 3pt,high}\left(1, {\bf 1}, \Delta t, \Delta t^\prime\right)
-\tilde{C}_{\rm 3pt,high}\left(1, {\bf 1}, \Delta t, \Delta t^\prime\right)
\nonumber \\
& & 
\hspace{10mm}
+ 
\frac{1}{N_{\rm src,high}}
\sum_{t_{\rm src}=1,3,\cdots}^{2N_{\rm src,high}-1}
\tilde{C}_{\rm 3pt,high} \left(t_{\rm src}, {\bf y}_{\rm src}(t_{\rm src}), \Delta t, \Delta t^\prime \right)
,
\end{eqnarray}
where ${\bf 1}$ denotes the origin of the lattice.
We use the stopping condition $|D\psi_{\rm pt}-b|\leq10^{-7}$
for $C_{\rm 3pt,high}$,
and a more relaxed one $10^{-2}$ 
for the approximated estimator $\tilde{C}_{\rm 3pt,high}$.
In this study, we average $\tilde{C}_{\rm 3pt,high}$ over $N_{\rm src,high}\!=\!24$
source points, namely one point per two time-slices.

Figure~\ref{fig:lmaama} demonstrates 
the improvement of the statistical accuracy of the axial charges 
by LMA and TSM.
We observe about a factor of five improvement by LMA
in the low-mode contribution to the isovector charge $g_A$. 
Then the statistical error of $g_A$ is largely dominated 
by that of the high-mode contribution,
and is reduced by a factor of about four by TSM.
Note that these gains are (only) slightly smaller than the ideal values, 
$\sqrt{N_{\rm src,low}}\!\sim\!7$ and $\sqrt{N_{\rm src,high}}\!\sim\!5$,
due to the correlation in each configuration.
We observe that 
LMA and TSM are less effective for the isoscalar and strange-quark charges,
which also show large statistical error.

\section{Numerical results}

\begin{figure}[t]
\begin{minipage}{0.5\hsize}
\begin{center}
\includegraphics[width=7.5cm]{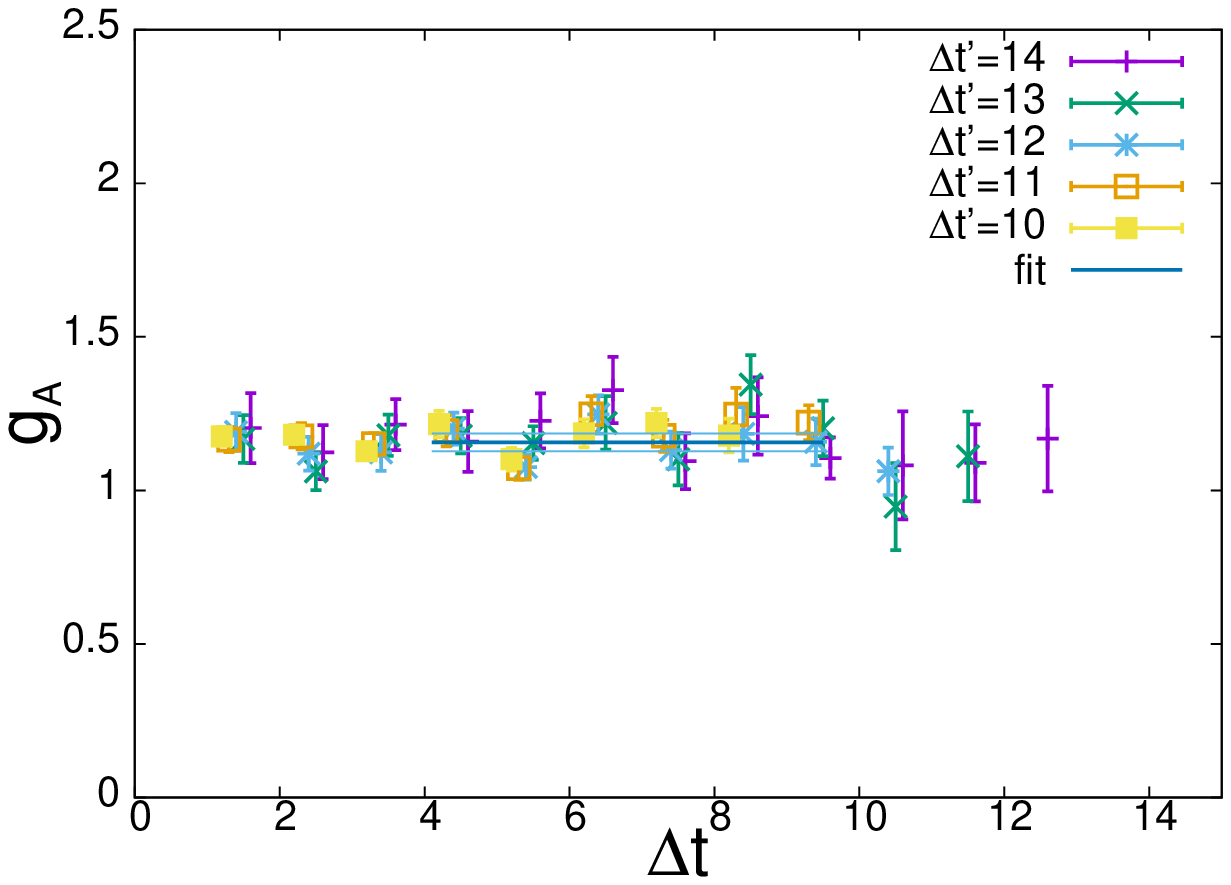}
\end{center}
\end{minipage}
\begin{minipage}{0.5\hsize}
\begin{center}
\includegraphics[width=7.5cm]{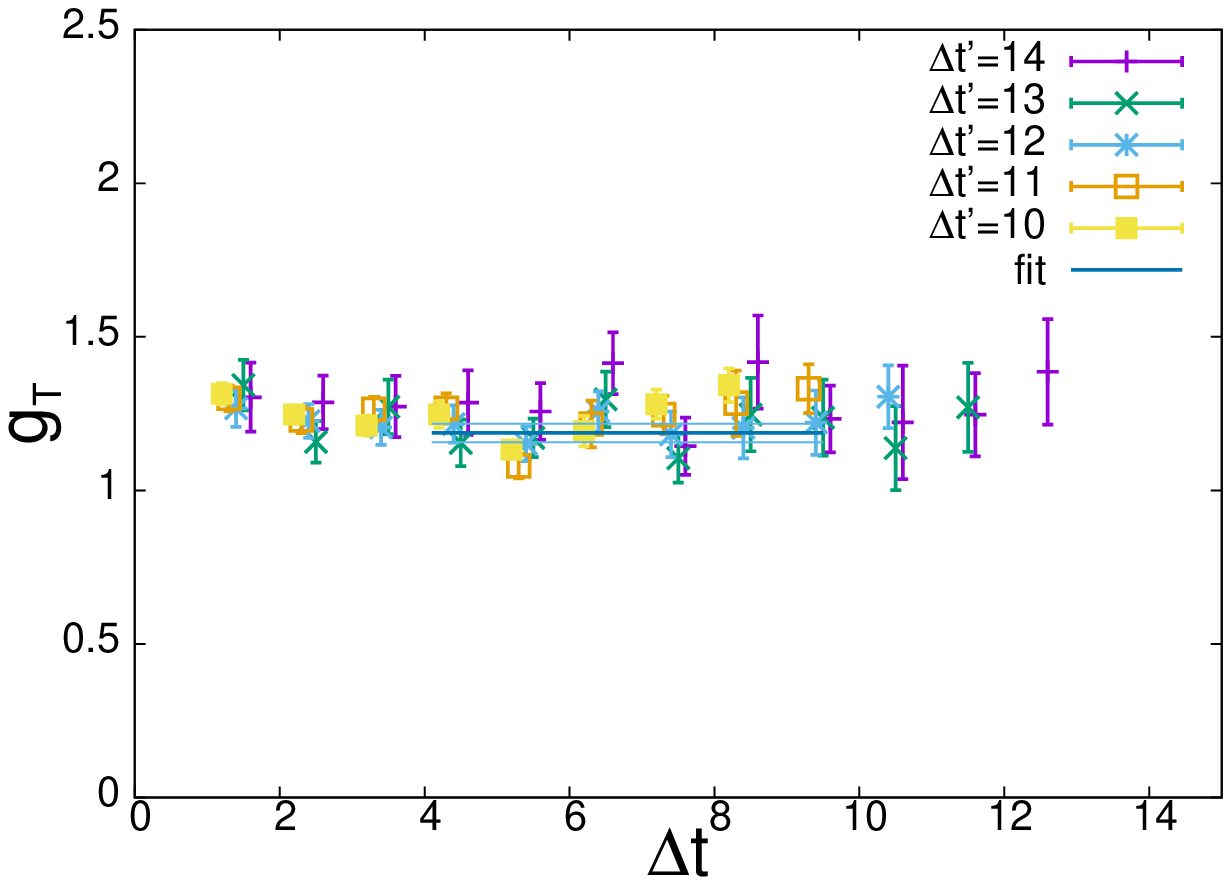}
\end{center}
\end{minipage}
\caption{\label{fig:isovector}
Effective value of 
isovector axial (left panel) and tensor (right panel) charges
at $m_\pi\!=\!540$~MeV.
We plot data with different values of $\Delta t^\prime$ by different 
symbols and the constant fit by solid lines.
}
\end{figure}

\begin{figure}[t]
\begin{minipage}{0.5\hsize}
\begin{center}
\includegraphics[width=7.5cm]{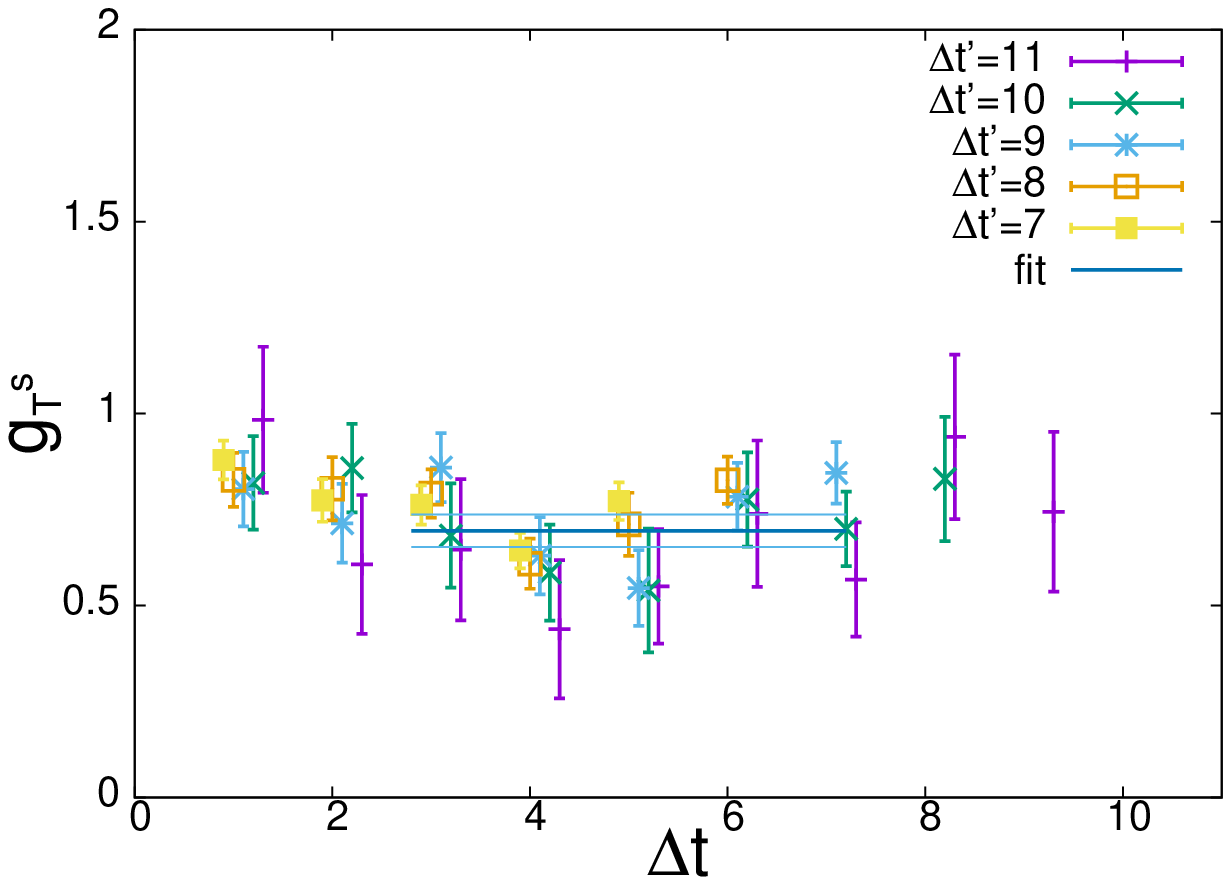}
\end{center}
\end{minipage}
\begin{minipage}{0.5\hsize}
\begin{center}
\includegraphics[width=7.5cm]{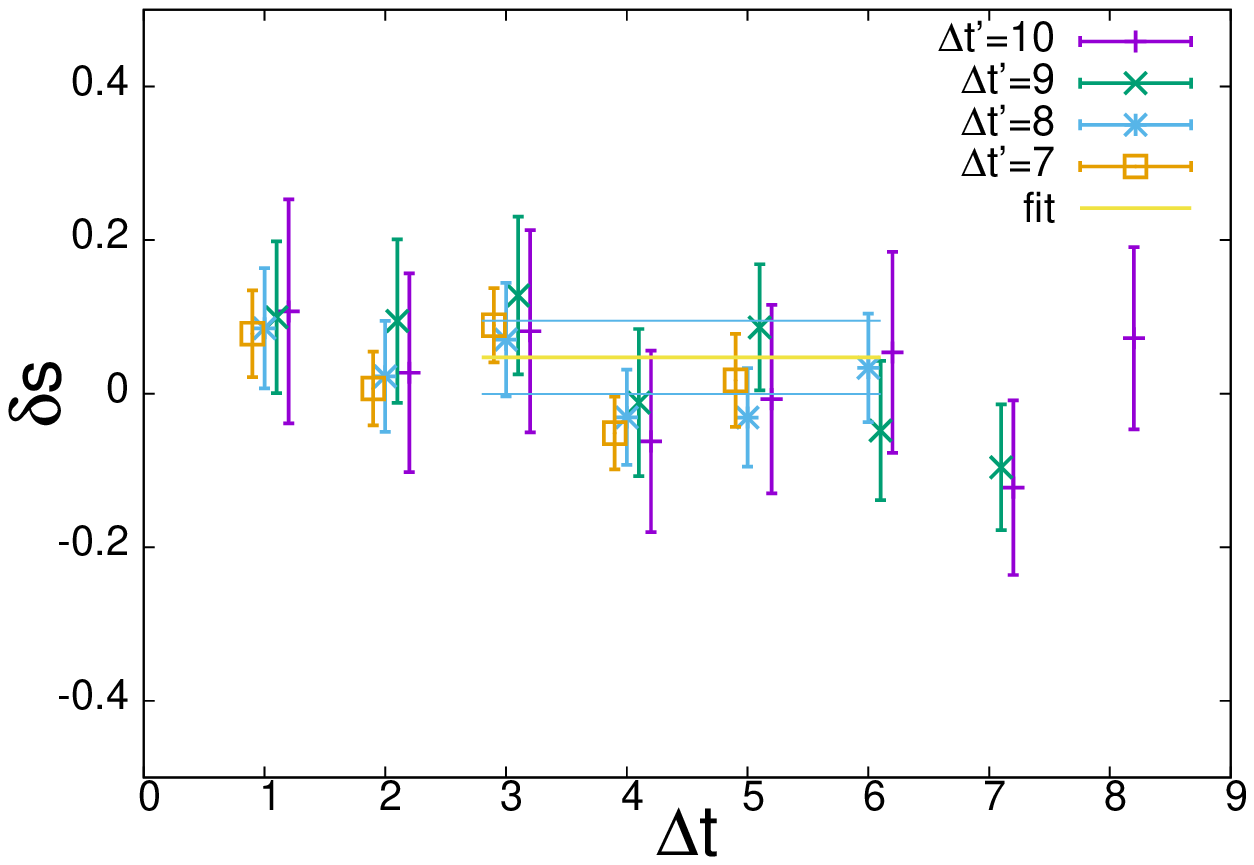}
\end{center}
\end{minipage}
\caption{\label{fig:isoscalar}
Effective value of isoscalar tensor charge (left panel)
and strange-quark tensor charge (right panel) at $m_\pi\!=\!540$~MeV.
}
\end{figure}

In Fig.~\ref{fig:isovector}, we plot the effective value of 
the isovector charges, $g_A$ and $g_T$,
obtained from the ratio~(\ref{eq:nucleoncharge}).
Data are stable against the choice of $\Delta t$
and $\Delta t^\prime$ suggesting that 
the excited state contamination is reasonably suppressed 
with our choice of the smeared nucleon operator.
We determine the charges by a constant fit to these data. 
The statistical error is typically 3\,\% for both $m_\pi\!=\!450$ and 540~MeV
with our simulation method using the all-mode averaging technique.

The effective value of the isoscalar tensor charge $g_T^s$
is plotted in the left panel of Fig.~\ref{fig:isoscalar}.
We observe that the isoscalar charges, $g_T^s$ and $g_A^s$, 
have larger statistical error, typically 10\,\%,
due to the presence of the noisy disconnected contribution.
On the other hand, the strange-quark charges, $\Delta s$ and $\delta s$,
consist solely of the disconnected contribution.
As shown in the right panel of Fig.~\ref{fig:isoscalar},
these charges are consistent with zero within the statistical error.

\begin{figure}[htbp]
\begin{minipage}{0.5\hsize}
\begin{center}
\includegraphics[width=7.5cm]{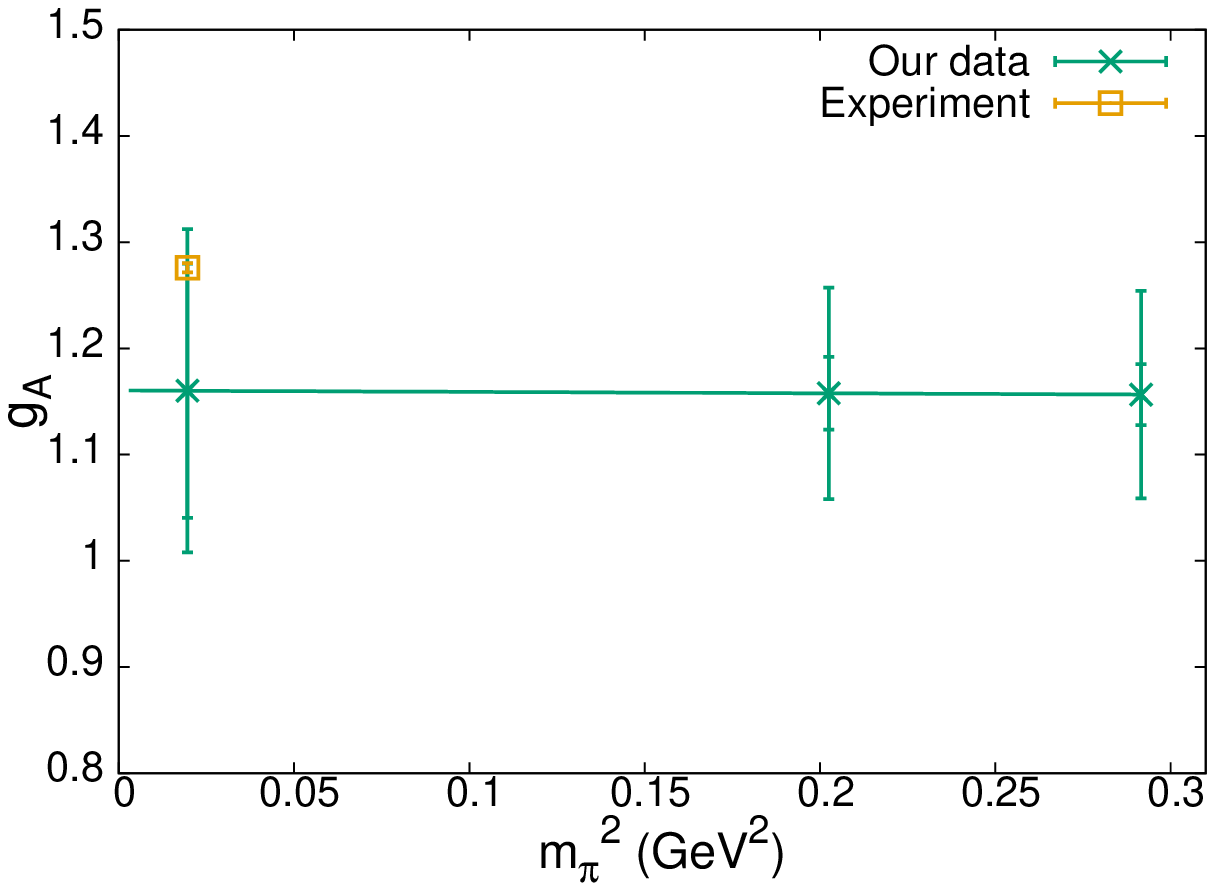}
\end{center}
\end{minipage}
\begin{minipage}{0.5\hsize}
\begin{center}
\includegraphics[width=7.5cm]{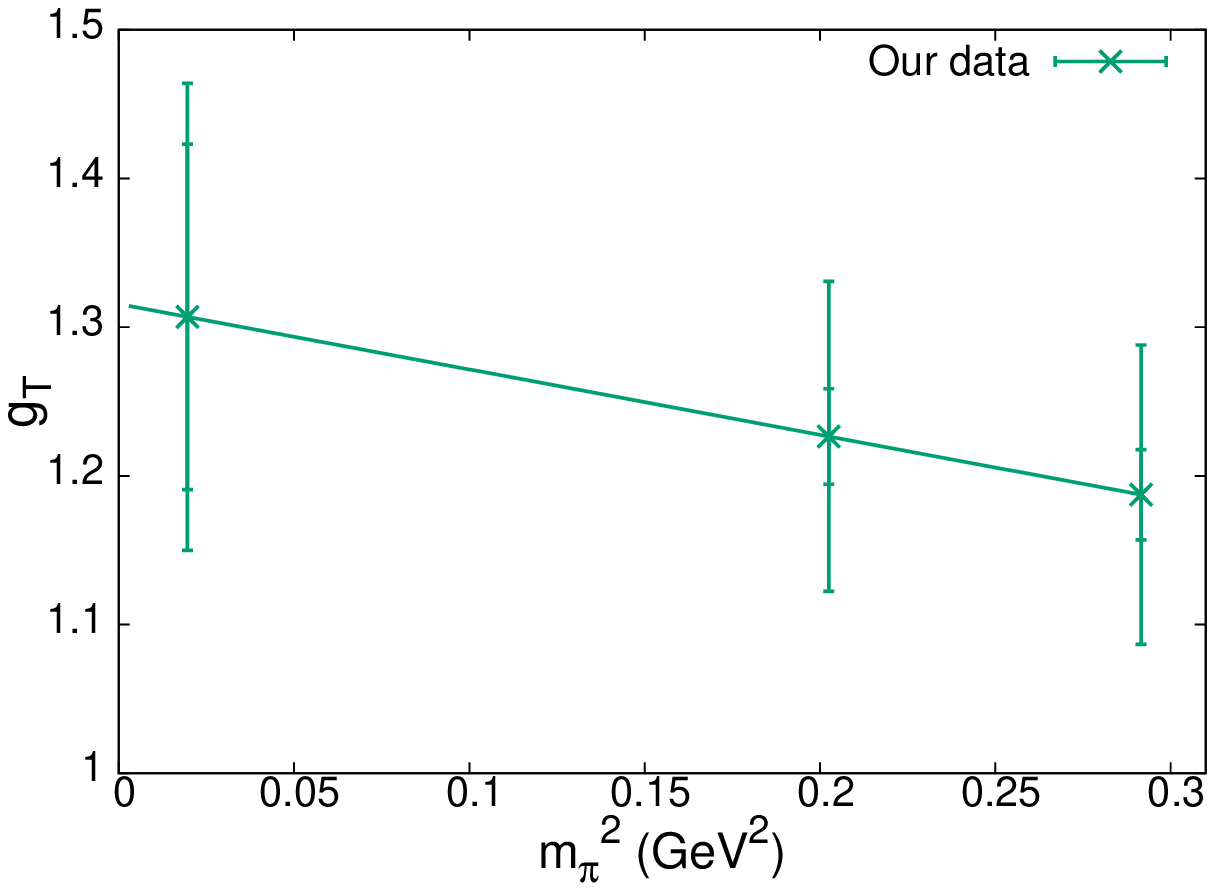}
\end{center}
\end{minipage}
\caption{\label{fig:extrapolation_isovector}
Chiral extrapolation of isovector charges to physical pion mass.
The left and right panels show the extrapolation for $g_A$ and $g_T$,
respectively. 
We also plot the experimental value~\cite{ucna} for $g_A$.
}
\end{figure}

Figure~\ref{fig:extrapolation_isovector} shows 
the chiral extrapolation of the isovector charges.
We observe their mild $m_\pi^2$ dependence, 
and the data are consistent with previous lattice studies of $g_A$~\cite{review:lat15}.
We employ a simple linear extrapolation in terms of $m_\pi^2$,
and obtain 
\begin{eqnarray}
g_A 
&=&
1.16(12)(9),
\hspace{5mm}
g_T
=
1.31(12)(11)
,
\end{eqnarray}
where the first error is statistical,
and the second is the discretization error estimated by
power counting $O((a\Lambda_{\rm QCD})^2)\!\sim\!8$\,\%
with $\Lambda_{\rm QCD}\!=\!500$~MeV.
These are consistent with 
previous lattice studies, $g_A\!\approx\!1.10$\,--\,1.25 
and $g_T\!\approx\!0.95$\,--\,1.15~\cite{review:lat15,review:cd15}.

\begin{figure}[b]
\begin{minipage}{0.5\hsize}
\begin{center}
\includegraphics[width=7.5cm]{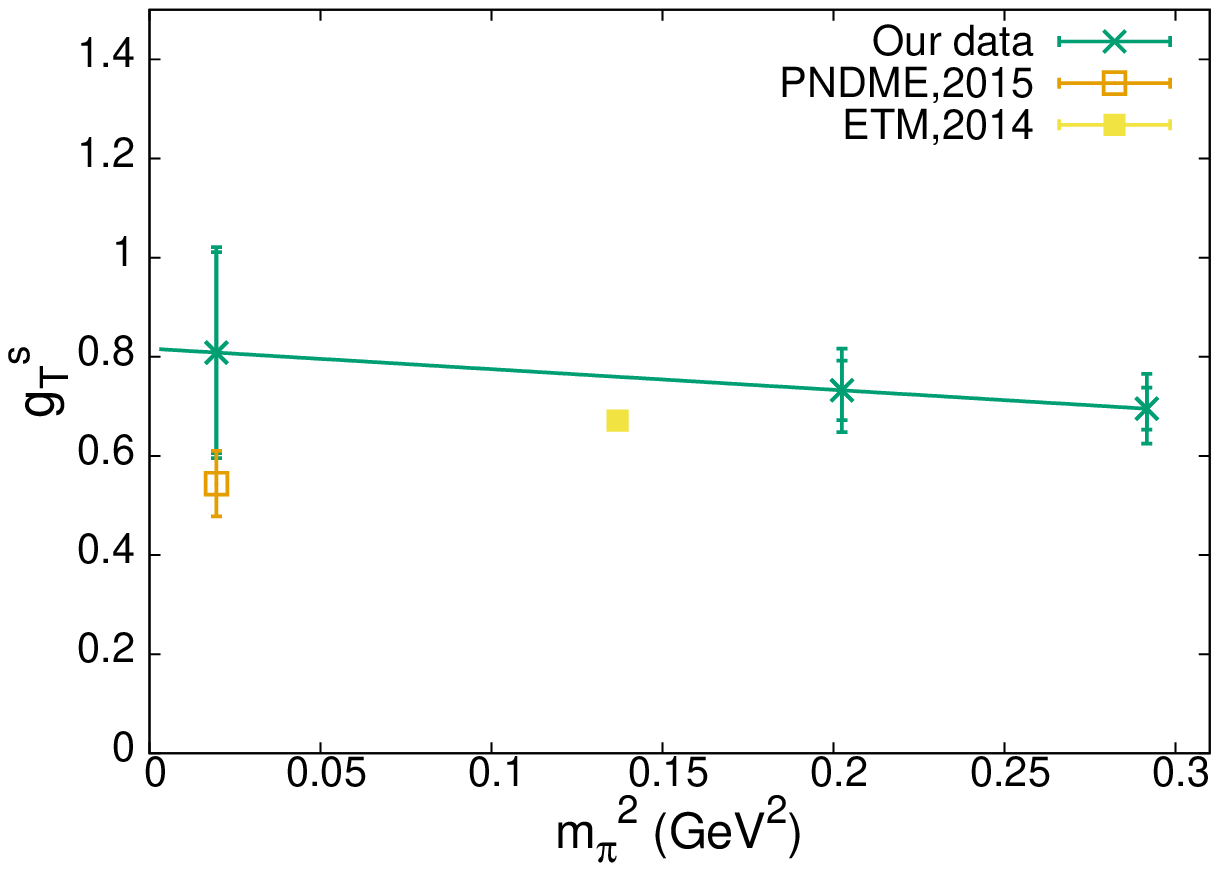}
\end{center}
\end{minipage}
\begin{minipage}{0.5\hsize}
\begin{center}
\includegraphics[width=7.5cm]{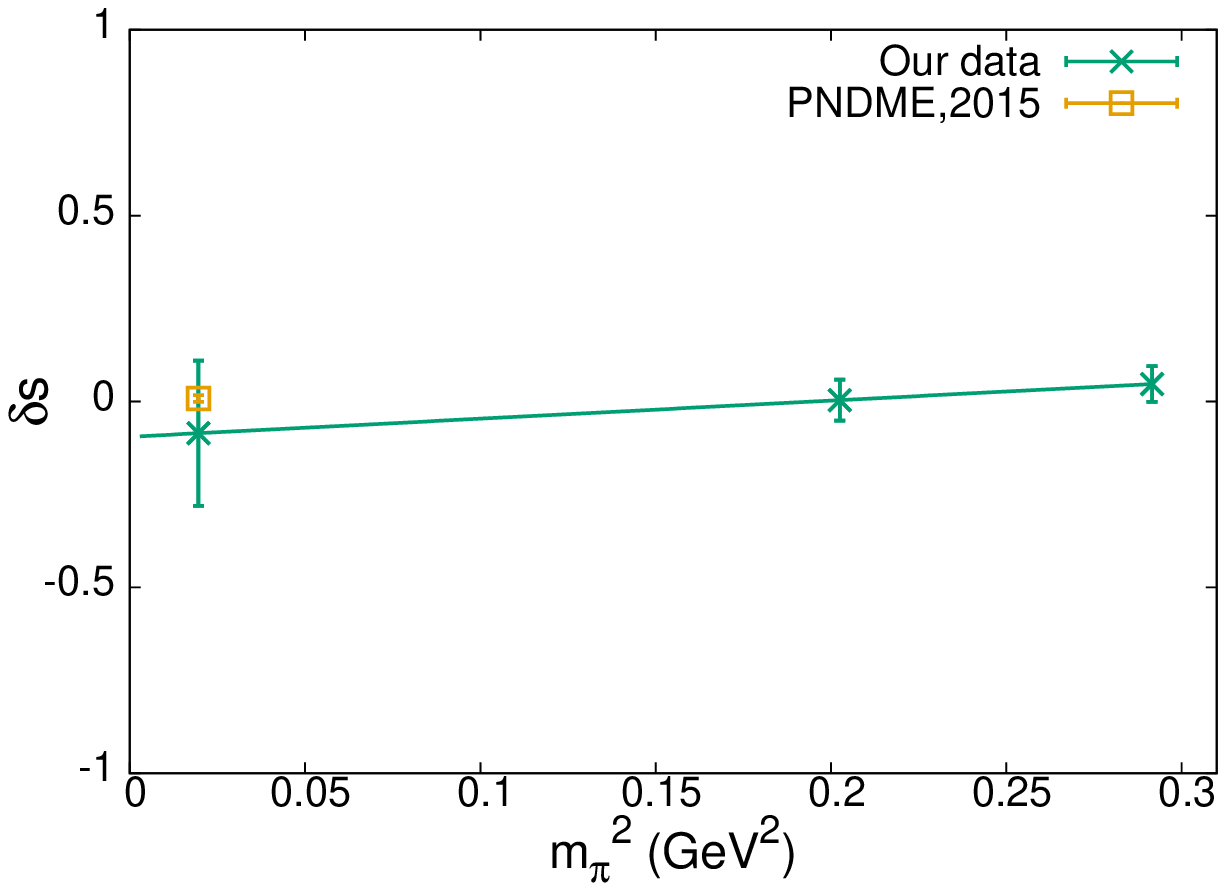}
\end{center}
\end{minipage}
\caption{\label{fig:extrapolation_isoscalar}
Chiral extrapolation of isoscalar (left panel) 
and strange-quark (right panel) tensor charges.
}
\end{figure}

As shown in the left panel of Fig.~\ref{fig:extrapolation_isoscalar},
we also observe a small $m_\pi^2$ dependence for the isoscalar charges
partly because of the larger statistical error 
due to the disconnected contribution.
A linear chiral extrapolation yields 
\begin{eqnarray}
g_A^s
&=&
0.63(18)(5),
\hspace{5mm}
g_T^s
=
0.81(20)(7).
\end{eqnarray}
With the present simulation set-up, these isoscalar charges are 
determined with $\lesssim\!30$\,\% accuracy at the physical point. 
Our results for the strange-quark charges are consistent with zero
at simulated $m_\pi$'s and hence at the physical point
\begin{eqnarray}
\Delta s
&=&
-0.11(18)(1),
\hspace{5mm}
\delta s
=
-0.09(20)(1)
\end{eqnarray}
as shown in the right panel of Fig.~\ref{fig:extrapolation_isoscalar}.
We note that small strange-quark charges have also been observed 
in recent studies~\cite{etm2,pndme}.

Within the present uncertainty,
our results for the axial charges, $g_A$ and $g_A^s$, are consistent 
with the experimental values~\cite{ucna,compass}.
The suppression compared to the simple quark model estimate,
$g_A^s\!=\!1$ and $g_A\!=\!5/3$, was argued by one of the authors
in Schwinger-Dyson analyses~\cite{yamanaka1,yamanaka2}.
For more precise comparison with experiment, 
however, we need a better control of the chiral extrapolation 
and discretization error.

\section{Summary}

We have reported on our calculation of 
the nucleon axial and tensor charges in lattice QCD 
with dynamical overlap fermions.
Disconnected nucleon correlation functions are calculated 
by using the all-to-all quark propagator.
We also employ the all-mode averaging technique,
namely LMA and TSM in this study, 
for which we demonstrate their efficiency for
both connected and disconnected functions.

Our preliminary results are consistent with 
previous lattice studies and experiments.
For more precise determination, 
we are testing different set-ups of LMA and TSM, 
for instance $N_{\rm src,low}$ and $N_{\rm src,high}$, 
to reduce the statistical error at simulated $m_\pi$'s.
Our on-going calculations at lighter $m_\pi$'s allow 
a controlled chiral extrapolation, and hence help 
improving the statistical accuracy at the physical point.
It is also important to reduce the discretization error 
especially for the isovector charges. 
Simulations on finer lattices with a different chiral fermion formulation
are also in progress~\cite{Noaki:Lat14}.
\vspace{2mm}

The numerical calculations were performed on Hitachi SR16000 at High Energy Accelerator Research Organization under a support of its Large Scale Simulation Program (No.~15/16-09), and on Hitachi SR16000 at Yukawa Institute of Theoretical Physics.
This work is supported in part by the Grant-in-Aid of the MEXT 
(No.~26247043 and 26400259),
RIKEN iTHES Project, RIKEN Special Postdoctoral Researcher program, 
MEXT SPIRE and JICFuS.


\begin{thebibliography}{99}

\bibitem{ohki}
H. Ohki {\it et al.} (JLQCD Collaboration), 
Phys. Rev. D {\bf 78}, 054502 (2008).

\bibitem{takeda}
K. Takeda {\it et al.} (JLQCD Collaboration),
Phys. Rev. D {\bf 83}, 114506 (2011).

\bibitem{ohki2}
H. Ohki {\it et al.} (JLQCD Collaboration),
Phys. Rev. D {\bf 87}, 034509 (2013).

\bibitem{exW}
H.~Fukaya {\it et al.} (JLQCD Collaboration),
Phys. Rev. D {\bf 74}, 094505 (2006).

\bibitem{noaki}
J.~Noaki {\it et al.} (JLQCD Collaboration), 
Phys. Rev. D {\bf 81}, 034502 (2010).

\bibitem{A2A:SESAM} 
G.S.~Bali {\it at al.} (SESAM Collaboration),  
Phys. Rev. D {\bf 71}, 114513 (2005).

\bibitem{A2A:TrinLat} 
J.~Foley {\it et al.} (TrinLat Collaboration),
Comput. Phys. Commun. {\bf 172}, 145 (2005).

\bibitem{noise} 
S.-J.~Dong and K.-F.~Liu,
Phys. Lett. B {\bf 328}, 130 (1994).

\bibitem{blumama}
T.~Blum, T.~Izubuchi and E.~Shintani, 
Phys. Rev. D {\bf 88}, 094503 (2013).

\bibitem{lma:ds}
T.~DeGrand and S.~Schaefer, 
Comput. Phys. Commun. {\bf 159}, 185 (2004).

\bibitem{lma:ghlww}
L.~Giusti, P.~Hernandez, M.~Laine, P.~Weisz and H.~Wittig,
JHEP {\bf 0404} 013 (2004).

\bibitem{tsm}
G.S.~Bali, S.~Collins and A.~Schafer, Comp. Phys. Com. {\bf 181}, 1570 (2010).

\bibitem{review:lat15}
J.~Zanotti, in these proceedings.

\bibitem{review:cd15}
M.~Constantinou, arXiv:1511.00214 [hep-lat].

\bibitem{etm2}
A.~Abdel-Rehim {\it et al.}, Phys. Rev. D {\bf 89}, 034501 (2014).

\bibitem{pndme}
T.~Bhattacharya {\it et al.}, Phys. Rev. D {\bf 89}, 094502 (2014).

\bibitem{ucna}
B.~Plaster {\it et al}. (UCNA Collaboration), Phys. Rev. C {\bf 86}, 055501 (2012).

\bibitem{compass}
M.G.~Alekseev {\it et al}. (COMPASS Collaboration), 
Phys. Lett. B {\bf 693}, 227 (2010).

\bibitem{yamanaka1}
N.~Yamanaka, T.M.~Doi, S.~Imai and H.~Suganuma,
Phys. Rev. D {\bf 88}, 074036 (2013).

\bibitem{yamanaka2}
N.~Yamanaka, S.~Imai, T.M.~Doi and H.~Suganuma,
PHys. Rev. D {\bf 89}, 074017 (2014).

\bibitem{Noaki:Lat14}
J.~Noaki {\it et al.} (JLQCD Collaboration),
PoS {\bf LATTICE2014}, 069 (2015).


\end{thebibliography}
\end{document}